%% file: main.tex
\documentclass[sigconf]{acmart}
\AtBeginDocument{%
  }

\copyrightyear{2026}
\acmYear{2026}
\setcopyright{cc}
\setcctype{by}
\acmConference[ICCAD '26]{IEEE/ACM International Conference on Computer-Aided Design}{November 08--12, 2026}{San Jose, CA, USA}
\acmBooktitle{IEEE/ACM International Conference on Computer-Aided Design (ICCAD '26), November 08--12, 2026, San Jose, CA, USA}
\acmDOI{10.1145/3831252.3834007}
\acmISBN{979-8-4007-2873-0/2026/11}

\usepackage{booktabs}
\usepackage{tabularx}
\usepackage{makecell}
\newcolumntype{R}{>{\raggedleft\arraybackslash}X}
\newcolumntype{C}{>{\centering\arraybackslash}X}
\usepackage{eso-pic}
\usepackage{tikz}

\AddToShipoutPictureFG*{%
  \begin{tikzpicture}[remember picture, overlay]
    \node[anchor=north, yshift=-0.4cm] at (current page.north) {%
      \parbox{0.9\textwidth}{\centering\small
        This paper is accepted for publication at the International Conference on Computer Aided Design (ICCAD 2026) {November 08--12,
  2026, San Jose, CA} %
      }%
    };
  \end{tikzpicture}%
}

\begin{document}

\title{SPARC: Automated Root-Cause Analysis of Pre-Silicon Power Side-Channel Leakage in the Processor Design Flow}

% \author{Andrija Ne\v{s}kovi\'{c}}

\author{Andrija Ne\v{s}kovi\'{c}}
\email{andrija.neskovic@uni-luebeck.de}
\affiliation{%
  \institution{Institute of Computer Engineering, Universität zu Lübeck}
  \city{Lübeck}
  \country{Germany}
}
\author{Christian Ewert}
\email{christian.ewert@uni-luebeck.de}
\affiliation{%
  \institution{Institute of Computer Engineering, Universität zu Lübeck}
  \city{Lübeck}
  \country{Germany}
}
\author{Mladen Berekovic}
\email{mladen.berekovic@uni-luebeck.de}
\affiliation{%
  \institution{Institute of Computer Engineering, Universität zu Lübeck}
  \city{Lübeck}
  \country{Germany}
}
\author{Saleh Mulhem}
\email{saleh.mulhem@uni-luebeck.de}
\affiliation{%
  \institution{Institute of Computer Engineering, Universität zu Lübeck}
  \city{Lübeck}
  \country{Germany}
}

\renewcommand{\shortauthors}{Ne\v{s}kovi\'{c} et al.}

\input{sections/00-abstract}




\maketitle

\input{sections/01-introduction}

\input{sections/02-background}
\input{sections/03-methodology}
\input{sections/04-experimental-setup-evaluation}

\input{sections/06-conclusion}

\begin{acks}
This work was partially funded by the German Federal Ministry of Research, Technology, and Space (BMFTR) through the project DI-OSVISE (16ME0958).
\end{acks}

\bibliographystyle{ACM-Reference-Format}
\bibliography{sample-base}

\end{document}

%% file: sections/00-abstract.tex
\begin{abstract}
Power‑Side‑Channel Leakage (PSCL) originates from architectural and micro‑architectural artifacts in a processor and poses a severe threat to the confidentiality of cryptographic software. Consequently, pre‑silicon PSCL evaluation is indispensable for secure hardware design. Existing frameworks are either limited by poor simulation scalability or fail to attribute leakage to the correct hardware signals and software instructions, thereby impeding a comprehensive root‑cause analysis. This paper presents SPARC, an automated framework for pre‑silicon PSCL evaluation and root‑cause analysis. SPARC leverages macro‑cell‑level Information Flow Tracking (IFT) augmented with enhanced shadow logic that tags switching activity originating from secret‑dependent data. By isolating this activity, SPARC applies statistical leakage tests to detect PSCL, while simultaneously attributing the leakage to specific hardware signals and mapping those signals to the corresponding software instructions. This approach thus delivers a full end‑to‑end leakage evaluation and root-cause analysis for both hardware and software. To demonstrate and validate SPARC, PSCL of multiple open‑source RISC‑V CPUs, encompassing 32‑bit and 64‑bit cores with both in‑order and out‑of‑order pipelines, is evaluated across a range of cryptographic workloads, including masked and unmasked AES and ML‑KEM (CRYSTALS-Kyber-512). SPARC recovers known leakage sources as a sanity check and identifies specific microarchitectural leakage sources, achieving an 8x per-trace simulation speedup over previously shown approaches on comparable designs. By enabling precise and scalable root‑cause analysis at the pre‑silicon stage, this work provides a practical framework to mitigate PSCL early in the design flow, thereby strengthening the security of future processors.
\end{abstract}

%% file: sections/01-introduction.tex
\section{Introduction}
\label{sec:intro}
Since the introduction of Power Side-Channel Attacks (PSCA)~\cite{kocher1999differential, brier2004correlation}, the notion of secure implementation has expanded beyond algorithmic cryptographic security to include implementation-specific vulnerabilities, such as Power Side-Channel Leakage (PSCL). 
Although software and hardware countermeasures like masking~\cite{gigerl2021secure} can mitigate PSCL, vulnerabilities at the architectural and microarchitectural levels of CPUs may compromise these protections~\cite{adhikary2026root, desmet2026masking}. 
Consequently, a precise pre‑silicon root‑cause analysis is essential to identify the exact hardware signals and software instructions that leak secret data, thereby enabling targeted countermeasure deployment.
Current hardware‑level root‑cause analysis techniques suffer from fundamental limitations. Their reliance on complete gate‑level netlists often results in runtimes exceeding 60 hours for designs with roughly 100k cells, rendering them impractical for most designs~\cite{yao2020architecture}. 
Recent software‑level approaches either ignore a large portion of switching activity, limiting scalability to complex CPU pipelines~\cite{liu2025telescope}, or focus solely on software behavior, leaving the underlying hardware unexplored~\cite{adhikary2025archer}.
To the best of our knowledge, no framework simultaneously offers scalability, secret‑sensitivity awareness, and joint hardware‑and‑software root‑cause attribution in a pre‑silicon context.
To bridge this gap, a novel root-cause analysis methodology is essential to rigorously address three fundamental questions:
\begin{itemize}
    \item[\textbf{RQ1}]: At which simulation cycles does the design leak?
    \item[\textbf{RQ2}]: Which signals contribute to the leakage?
    \item[\textbf{RQ3}]: Which software instructions cause the leakage?
\end{itemize}
Collectively, these features enable efficient pre-silicon root-cause analysis that is scalable, secret-sensitive, and actionable for both hardware designers and software developers.

\subsection{Contribution}
In this paper, we present SPARC \footnote{Source code available: \url{https://github.com/iti-luebeck/sparc-sca}}, an automated framework for pre-silicon PSCL evaluation and root-cause analysis of cryptographic software executing on CPUs. The primary innovation is to deploy Information Flow Tracking (IFT) at the macro-cell level for PSCL evaluation. This enables scalable identification of leakage sources across both hardware signals and software instructions.
%In this paper, we present SPARC, an automated framework for pre-silicon PSCL evaluation and root-cause analysis of cryptographic software executing on CPUs.
%The primary innovation is the integration of macro cell-level Information Flow Tracking (IFT) with dedicated leakage estimation logic, which enables scalable identification of leakage sources across both hardware signals and software instructions.
%We present an automated framework for pre-silicon PSCL evaluation and root-cause analysis of cryptographic software executing on CPUs. The primary innovation is the integration of macro cell-level Information Flow Tracking (IFT) with dedicated leakage estimation logic, which enables scalable identification of leakage sources across both hardware signals and software instructions.
\begin{figure}
    \centering
    \includegraphics[width=1\linewidth]{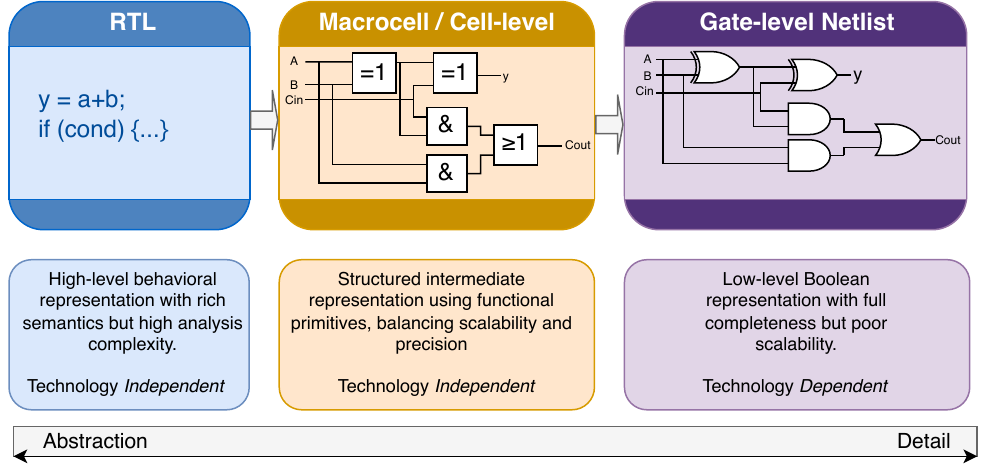}
    \caption{Comparison of hardware abstraction levels}
    \Description{Three side-by-side panels showing the same full-adder circuit at increasing levels of detail. Left, RTL: behavioral HDL code with an addition and a conditional, described as technology-independent with rich semantics but high analysis complexity. Center, macro-cell level: a structured schematic of parameterized XOR, AND, and OR primitives with inputs A, B, and carry-in and outputs y and carry-out, described as technology-independent and balancing scalability with precision. Right, gate-level netlist: the same function as discrete Boolean gates, described as technology-dependent, fully complete, but poorly scalable. A horizontal arrow beneath the panels runs from abstraction on the left to detail on the right.}
    \label{fig:abstraction_levels}
    
\end{figure}
The macrocell abstraction level serves as an intermediate representation between RTL and gate-level netlists, as illustrated in Fig.~\ref{fig:abstraction_levels}. 
In this abstraction, the targeted design is represented as graphs of parameterized functional primitives such as adders, multiplexers, and comparators. 
This abstraction preserves the semantic structure of the design while avoiding the complexity of HDL constructs and the scalability challenges of full gate-level representations. 
As previously demonstrated in~\cite{solt2022cellift}, instrumenting designs at this level with IFT logic enables tracking of secret propagation in complex CPUs and full SoCs. 
Building upon this, SPARC monitors the flow of secret data through the hardware during cryptographic software execution, tracking only activity that is demonstrably secret-dependent. 
This focus on secret‑sensitive switching directly mitigates the scalability limitations that have plagued previous PSCL evaluation methods.
The key contributions of SPARC are as follows:
\begin{itemize}
    \item \textbf{Cycle‑accurate simulation environment}: A dedicated transformation pass (realized as Yosys pass~\cite{wolf2013yosys}) that instruments RTL designs with macro-cell level IFT and PSCL estimation logic. This pass is integrated into Verilator~\cite{verilator} for executing cryptographic software and collecting secret-dependent leakage traces.
    \item \textbf{Automated dual root‑cause analysis}: Our leakage evaluation pipeline applies statistical tests to detect leaking cycles, then performs a two‑phase attribution:
        \begin{itemize}
            \item Pinpointing leaky hardware signals within the design.
            \item Mapping those signals back to the executing software instructions.
        \end{itemize}
    The proposed leakage-estimation logic extends the IFT instrumentation to model both value-based and transition-based leakage, quantified with Hamming‑Weight (HW) and Hamming ‑Distance (HD) metrics. This yields a fast, scalable pre‑silicon evaluation flow that circumvents the runtime burden of gate‑level correlation analysis while delivering finer attribution than typical RTL‑level approaches.
    
    \item \textbf{Empirical validation}: We evaluate both masked and unmasked implementations of AES 128  and ML‑KEM (CRYSTALS Kyber-512) on a spectrum of open‑source RISC‑V cores (32‑bit and 64‑bit, in‑order and out‑of‑order pipelines). The results demonstrate SPARC’s effectiveness in detecting and pinpointing side‑channel leakage across diverse processor architectures.
\end{itemize}

%% file: sections/02-background.tex
\section{Background and Related Work}

Addressing PSC attacks \cite{kocher1999differential, brier2004correlation} in modern processors requires understanding three fundamental challenges. 
First, precisely characterizing how and where PSCL arises in hardware and software. 
Second, tracking the propagation of secrets through complex digital designs in a scalable manner. 
Third, attributing leakage sources to both specific hardware signals and the software instructions that trigger them, enabling feedback for designers and developers. In this section, we establish the necessary background across these three dimensions and identify the research gap that motivates our proposed approach.

\subsection{Power Side Channel Leakage}
\label{section:power-sca}

Cryptographic implementations are susceptible to side-channel attacks, in which sensitive data can be extracted by examining a device's physical characteristics during operation. 
Exploitable power characteristics arising from input-dependent switching in CMOS technology enable adversarial side-channel attacks \cite{standaert2010introduction-SCA}, rendering mathematically secure algorithms vulnerable in practice.
The amount of current that a CMOS circuit draws at any given moment is greatly influenced by how often its transistors turn \textit{on} and \textit{off}. 
This activity level is determined by the data being handled within the circuit, and is referred to as \emph{dynamic power} consumption. 
Attackers exploit the relationship between dynamic power consumption and data values to reveal secret keys with PSC attacks. Each time a gate switches, its load capacitance is charged or discharged. In general, we can model this dynamic power consumption as:
\begin{equation}
\label{eq:power_dynamic}
P_{\text{dyn}} = \alpha \times C_L \times V_{dd}^2 \times f
\end{equation}

where $\alpha$ is the activity factor (data-dependent), $C_L$ is the load capacitance, $V_{dd}$ is the supply voltage, and $f$is the clock frequency. Because $\alpha$ depends on the processed data, the power drawn by a circuit reflects the data values \cite{jan2003digital}, providing the basis for PSC attacks.

To reason about PSCL analytically, the data-dependent activity factor $\alpha$ must be instantiated as a concrete leakage model that approximates how a circuit's power consumption relates to the data it processes. Two leakage models are well established in the literature.
First, the~\emph{Hamming Weight} (HW) model assumes that power consumption is proportional to the number of ones in the data word being processed.

This model is appropriate when the dominant contribution to switching activity comes from the absolute value of a signal: for instance, when a register is written from a constant or zero state.
Second, the~\emph{Hamming Distance} (HD) model instead captures transition-based leakage, where power consumption depends on how many bits change between two consecutive states.
This model is more appropriate for sequential elements such as registers and flip-flops, where each clock cycle causes a transition from a previous state to a new one, and the switching activity is determined by the difference between them. Both models serve as the basis for the leakage estimation logic introduced in this work.
Detecting whether a device's power consumption leaks secret-dependent information is framed as a statistical hypothesis test that relies on one of the previous leakage models. 
The standard methodology is \emph{Test Vector Leakage Assessment} (TVLA)~\cite{gilbert2011tvla}, which uses Welch's $t$-test. 
When $|t|$ value exceeds a threshold (conventionally 4.5 \cite{gilbert2011tvla}), it indicates that the device leaks information about the processed data.

\subsection{Information Flow Tracking}
\label{section:IFT}
Hardware Information Flow tracking enables the tracking of information propagation through an integrated circuit~\cite{hu2021hardware}. 
Static IFT is performed by analyzing all possible inputs to provide formal security guarantees without running a simulation. 
In practice, this scales poorly with growing design sizes, making the approach impractical for large designs ~\cite{hu2021hardware}. 
Dynamic IFT instead propagates \textit{taint} labels during simulation with chosen inputs, delivering a more scalable approach but offering no guarantees about unexercised paths. 
IFT can be performed at various abstraction levels, ranging from the software and architecture level down to the RTL~\cite{ardeshiricham2017register}, macro-cell level~\cite{solt2022cellift}, and gate-level netlist (referred to as Gate Level Information Flow Tracking (GLIFT) \cite{tiwari2009complete}). 
Lower abstraction levels capture more precise information flows, with more detailed signal interactions and timing accuracy at the cost of complexity. 
Among these, CellIFT~\cite{solt2022cellift} operates at the macro-cell level by replacing each functional primitive in the RTL netlist with a taint-propagating counterpart, propagating taint labels through parameterized cells such as adders, multiplexers, and comparators without decomposing them into individual gates. 
This retains semantic structure while keeping the shadow logic overhead manageable, making it well-suited as the foundation for scalable IFT in complex designs such as CPUs.

IFT was designed to track explicit and implicit information flows through the functional logic of a design, which manifests through data- and control-flow dependencies. 
Most of the existing work, therefore, covers classical security assertions and microarchitectural side channels, making no claims about PSCL evaluation~\cite{tiwari2009complete, ardeshiricham2017register, solt2022cellift}. 
To bridge this gap, Hamming Distance and Hamming Weight have been deployed to model the PSCL. 
In~\cite{zhang2021psc}, an automated approach to model PSCL at RTL using IFT was introduced. 
It built an RTL IFT engine using commercial EDA tools to trace taint propagation from key and plaintext inputs through cryptographic engines. 
Employing both formal assertions based on HW and HD, as well as t-test analysis (for masked designs), this approach evaluates the PSCL of AES, Simon, and Present block cipher hardware engines. 
However, it has been validated only on relatively small cryptographic cores and incurs substantial simulation overhead, indicating a runtime of 27 hours for the bit-serialized Simon t-test evaluation~\cite{zhang2021psc}. 
Consequently, this approach may not scale well to larger designs with longer simulation run time (such as CPUs or whole SoCs). 
Security evaluations targeting PSC were also deployed using GLIFT~\cite{wang2025pre}. 
It deploys Shannon entropy as a quantitative leakage metric to model the relation between taint label distributions and power consumption characteristics. 
While GLIFT evaluation strategy reduces the number of required simulations compared to exhaustive approaches, it still suffers from overall scalability issues.
Being implemented using GLIFT, the shadow logic doubles the design size, and the number of required simulations per key segment grows exponentially with the key segment size (8 bits in their experiments with cipher hardware engines)~\cite{wang2025pre}. 
Therefore, this approach is not practical for CPU or SoC evaluations.

\subsection{PSCL Root Cause Analysis}
\label{section:pre-silicone-PSCL_root_cause}
Several pre-silicon PSCL evaluation techniques have been proposed~\cite{he2019rtl, nahiyan2020script, gigerl2021secure, nevskovic2025sca, adhikary2025archer, liu2025telescope}. 
These approaches simulate the design at the ISA, RTL, or gate level, extract switching activity, or estimate dynamic power to generate execution traces that can be evaluated using well-known post-silicon techniques. Frameworks such as RTL-PSC \cite{he2019rtl}, SCRIPT \cite{nahiyan2020script}, and PSC-TG \cite{zhang2021psc} demonstrated the feasibility of PSCL assessment in pre-silicon. However, they offer little insight into which specific design elements are responsible for the leakage occurring. The need to explain leakage motivated the development of root-cause analysis techniques for PSCL.

To evaluate cryptographic software implementations on RISC-V processors, Archer \cite{adhikary2025archer} went beyond leakage detection to identify and explain instruction-level sources of leakage. This enables the evaluation of the impact of compiler decisions, such as instruction scheduling and register allocation, on PSCL in cryptographic software implementations. However, Archer operates at the ISA-simulation level and does not extend its analysis to the underlying hardware design, leaving the hardware sources of leakage unattributed.
Extending the root-cause analysis to hardware designs requires localizing leakage sources within a netlist. 
Architecture Correlation Analysis (ACA) was proposed in~\cite{yao2020architecture}. 
It ranks every cell in a gate-level netlist by its contribution using the Leakage Impact Factor (LIF), derived from the correlation between each cell's switching activity and a leakage model. 
In their experiments targeting an AES engine and an SoC with a Leon3 processor (approx. 100k cells), the authors show that only a small subset of gates contributes to leakage. 
While accurate, the 60-hour runtime for the relatively small design highlights the practical constraints of gate-level evaluation. 
Additionally, this approach provides no insight into the root cause analysis of software instructions.
In~\cite{liu2025telescope}, the authors proposed Telescope, a top-down hierarchical framework for PSCL evaluation at the architecture, microarchitecture, and gate levels, using toggle-count-based power estimation with TVLA. 
The root cause analysis in this framework can trace the PSCL back to individual software instructions. 
Nevertheless, its use of static timing analysis for signal path tracing, combined with the absence of information-flow tracking, requires considering all switching activity in the design rather than only secret-dependent signals, limiting scalability for complex CPU pipelines.

An orthogonal line of work formally verifies the absence of leakage: CoCo~\cite{gigerl2021coco} co-verifies masked software on a concrete CPU netlist, and Power Contracts~\cite{bloem2022power} establish provably complete leakage models. Both offer guarantees SPARC does not target, at high cost (up to 35 h per contract for Ibex, repeated after every hardware change); they suit certification of finalized designs, whereas SPARC targets fast assess–modify–reassess iterations during design.

\subsection{Motivation}

The highlighted state‑of‑the‑art techniques suffer from a steep simulation cost as their run‑time grows exponentially, and they do not provide a mechanism for precise root‑cause attribution that spans both hardware signals and software instructions.

Combining dynamic information-flow tracking with activity-based leakage estimation at the macro-cell level could overcome these shortcomings.  
Isolating secret-dependent switching activity through taint propagation enables scalable pre-silicon evaluation of complex CPU designs while supporting precise root-cause attribution across both hardware signals and software instructions, a capability no existing framework provides jointly.
Therefore, we propose SPARC to overcome the scalability and precision limitations of previous approaches to root-cause analysis of PSCL.
It proposes a new, dedicated leakage-estimation logic based on HW and HD models, alongside macro-cell-level IFT. This solution enables scalable, secret-sensitive PSCL evaluation and root-cause analysis directly on CPU designs.

%% file: sections/03-methodology.tex
\section{SPARC: Macro Cell Information Flow Tracking for PSCL}
\begin{figure}
    \centering
    \includegraphics[width=1\linewidth]{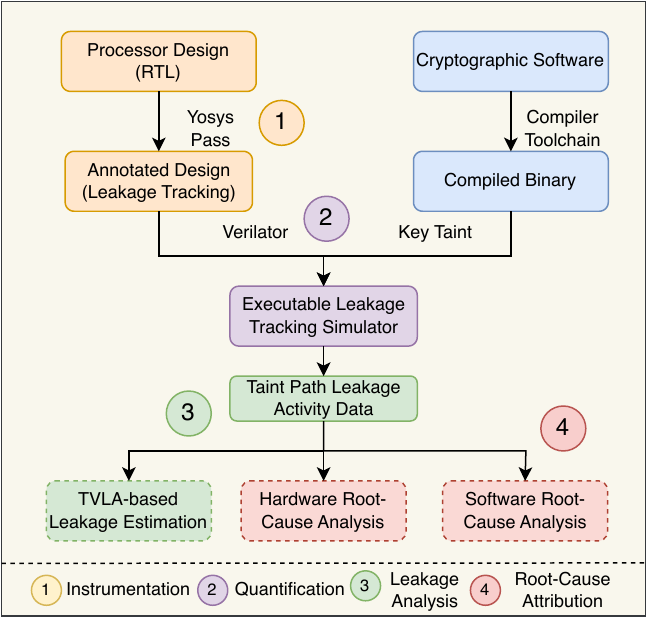}
    \caption{SPARC: Automated Framework Overview}
    \Description{Diagram showing the automated framework overview.}
    \Description{Dataflow diagram of the SPARC framework with two inputs at the top. The processor RTL design passes through a Yosys instrumentation pass to become an annotated design with leakage tracking; in parallel, cryptographic software passes through a compiler toolchain to become a compiled binary. Both feed into Verilator, together with the key taint specification, producing an executable leakage-tracking simulator that emits taint path leakage activity data. This data branches into three outputs: TVLA-based leakage estimation, hardware root-cause analysis, and software root-cause analysis. Numbered markers label the four phases: instrumentation, quantification, leakage analysis, and root-cause attribution.}
    \label{fig:framework}
\end{figure}

SPARC combines dynamic IFT with quantitative power modeling, aiming to enable pre-silicon side-channel leakage detection and root-cause analysis directly from simulation. 
SPARC is structured in four distinct phases (illustrated in Fig. \ref{fig:framework}:
(1) Instrumentation: Adapt a cell-level IFT pass to generate leakage indicators for each signal. 
(2) Quantification: Aggregate switching activity into scalar measurements representing leakage per cycle during simulation. 
(3) Evaluation: Analyze the collected data using statistical hypothesis testing and detailed attribution analysis to identify the cycles during which leakage occurs. 
(4) Root-cause analysis: Provide both hardware and software root-cause analysis.

\subsection{Instrumentation}
\label{section:instrumentation}
A signal in a synchronous design $D$ exhibits key-dependent power leakage if its switching activity differs statistically depending on computations with distinct keys $K_0$ and $K_1$ for the same plaintext. The signal $s$ leaks information depending on the dynamic power consumption $P_{dyn}$ if:
\begin{equation}
\label{eq:transition_leakage}
    P(\text{transitions}(s, t) \mid K_0) \neq P(\text{transitions}(s, t) \mid K_1)
\end{equation}

Two conditions must be satisfied: the signal must be tainted, meaning it is transitively influenced by $K$, and it must transition in a key-dependent way. A tainted signal that holds the same value across keys does not dissipate differential power, whereas an untainted transition carries no secret information.

To detect signals that satisfy both conditions simultaneously, cell-level IFT pass adds per-signal taint shadows $s_{t0}$ to track reachability from a designated secret input, along with a third signal layer, the leakage shadow $s_{leak0}$, defined as:
\begin{equation}
\label{eq:leakage_shadow}
    s_{\text{leak0}}[i][t] = (s[i][t] \oplus s[i][t-1]) \wedge s_{t0}[i][t]
\end{equation}

The XOR term identifies a transition since the previous clock cycle, while the AND operation with the taint shadow enforces key dependency. The result is a per-bit wire that is high precisely when bit $i$
of signal $s$ both transitions and is tainted in the current cycle. For combinatorial nodes, where transitions occur within rather than across clock cycles, a conservative model $s_{\text{leak0}} = s_{t0}$ is applied. Any input change observable at the output is considered as leakage propagation.
\subsection{Macro Cell-Level Leakage Quantification}
\label{section:leakage_quantification}
The instrumentation stage produces a leakage shadow for every tainted signal in the design. It generates logic to aggregate per-signal leakage indicators into two scalar counters per simulation cycle, thereby modeling dynamic power consumption.
The \textbf{Hamming Weight (HW) counter} measures how many tainted bits hold a logic-1 value in the current cycle, modeling charge-dependent power on tainted nodes.
The \textbf{Hamming Distance (HD) counter} measures how many tainted bits transitioned since the previous cycle. For example, if five flip-flops of a tainted 32-bit register toggle in a cycle, the HD counter accumulates 5, while the HW counter reflects the register's post-transition population count; summed over all tainted signals in scope, these counters form a cycle-accurate proxy for key-dependent dynamic power. The configurable aggregation scope allows counters to target a single module or to be instantiated across multiple sub-modules simultaneously, enabling comparison of leakage contributions from different pipeline stages, memory interfaces, or bus controllers within a single simulation run.

\subsection{Overall Leakage Evaluation}
\label{section:leakage_evaluation_methodology}
The aggregation counters produce a single HW and HD value per simulation cycle per trace. To determine whether these values carry key-dependent information, we require a statistical test that can distinguish genuine leakage from random variation across multiple simulation runs. We deploy TVLA \cite{gilbert2011tvla}, which applies Welch's t-test on a per-cycle basis across two groups of traces that differ only in the secret key used.

\subsubsection{Trace Generation} For each evaluation, we generate $N$ match\-ed pairs of simulation traces. Group $A$ executes under key $K_0$; group $B$ executes under key $K_1$. Within each pair, the plaintext is randomly sampled but held identical across both groups, and masking randomness (if applicable) is similarly re-sampled per trace but matched between groups. This matched-pair design eliminates plaintext-dependent artifacts and ensures that any statistical difference between groups is attributable exclusively to the key.
\subsubsection{Per-Cycle Hypothesis Testing} For each simulation cycle $t$, the HW (or HD) counter values across all traces in each group form two distributions. We apply Welch's t-test to these distributions:

\begin{equation}
t[t] = \frac{\bar{x}_0[t] - \bar{x}_1[t]}{\sqrt{\dfrac{s_0^2[t]}{N_0} + \dfrac{s_1^2[t]}{N_1}}}
\label{eq:tvla}
\end{equation}
where $\bar{x}_i[t]$, $s_i^2[t]$, and $N_i$ denote the sample mean, variance, and trace count for group $i$ at cycle $t$.

\subsubsection{Decision Criterion} A cycle is classified as leaking if $|t| > 4.5$, the threshold specified by the TVLA methodology~\cite{gilbert2011tvla}.

The outcome is a t-statistic time series $\{t[n]\}_{n=0}^{T}$ that provides both a binary pass/fail verdict and temporal localization of leakage occurrences within the execution. Cycles exceeding the threshold are grouped into contiguous leakage intervals, each characterized by its peak $|t|$ value and the corresponding simulation cycle. These intervals serve as entry points for the root-cause analysis stage, which identifies the specific signals and instructions responsible for the detected leakage.

\subsection{Root-Cause Analysis}
The TVLA evaluation identifies \textit{when} leakage occurs but not \textit{what} causes it. A t-statistic exceeding the threshold at a given cycle indicates that the aggregate taint-filtered switching activity is key-dependent, but does not reveal which specific signals carry the leaking information or which software instructions are executing at that moment. The root-cause analysis stage addresses this by attributing each detected leakage interval to specific hardware signals and software locations through two complementary analyses.
\subsubsection{Signal-Level Attribution} For each leakage interval identified by the TVLA stage, we perform a targeted comparison of simulation waveforms between the two key groups. Rather than examining the aggregate counters, this step inspects individual leakage shadow wires to determine which signals exhibit the largest key-dependent divergence. For each signal $s$ within a leakage window $W$, we compute the cumulative key-dependent Hamming weight difference:
\begin{equation}
\Delta_s = \sum_{t \in W} \text{HW}\bigl(s_{K_0}[t] \oplus s_{K_1}[t]\bigr)
\label{eq:signal-attr}
\end{equation}
where $s_{K_0}[t]$ and $s_{K_1}[t]$ denote the value of signal $s$ at cycle $t$ under keys $K_0$ and $K_1$ respectively. Signals are ranked by $\Delta_s$, with the highest-ranked signal identified as the dominant leakage carrier for that interval.
Because the leakage shadow instrumentation already filters for taint reachability, this ranking operates only over signals that are transitively influenced by the secret.

\subsubsection{Instruction-level Attribution} Concurrently, a second analysis maps each leakage interval to the software instruction executing at that point. The program counter value is recorded at every simulation cycle, along with the leakage counters. For each cycle within a leakage interval, the corresponding program counter is resolved to a source-level function and line number. Consecutive leaky cycles are grouped into contiguous intervals of the t-statistic time series. Each interval is characterized by its peak $|t|$ value, the dominant leaking signal, and the associated instruction. This process yields a direct mapping from each leakage peak to the executed software instruction.

The signal-level and instruction-level results are combined to produce a complete diagnostic for each leakage interval: the hardware signal carrying the key-dependent information, the software instruction that triggered it, and the statistical strength of the leakage. This combined output delivers an automated root-cause analysis and provides actionable information for both hardware designers (which pipeline register, bus, or memory structure to protect) and software developers (which cryptographic operation or memory access pattern to restructure).

%% file: sections/04-experimental-setup-evaluation.tex
\section{Implementation \& Evaluation}
This section details the experimental environment established to evaluate the proposed leakage analysis methodology. To promote reproducibility, the entire implementation pipeline is outlined. Comprehensive descriptions of SPARC's toolchain for automated hardware instrumentation are included, together with specifications of the three RISC-V processor targets and definitions of the execution parameters for the cryptographic workloads. RQ1 through RQ3, defined in Section \ref{sec:intro}, are addressed in Section \ref{sec:evaluation}.
\subsection{Experimental Setup}
In the following, we describe the practical implementation of the previously outlined methodology, including details on the toolchain and implementation, an introduction to the three RISC-V processors selected for evaluation, and an outline of the cryptographic workload and scenario generation strategy.
\subsubsection{Toolchain and Implementation}
SPARC implements dedicated leakage-estimation logic based on HW and HD models, along with macro-cell IFT, to enable automated leakage tracking and quantification. The instrumentation operates as a synthesis pass within the open-source Yosys \cite{wolf2013yosys} synthesis suite. Given a processor and its configuration, the pass processes an RTL design in a bottom-up topological order, instrumenting each module with taint shadows, leakage shadows, and aggregation counters as specified in the Section \ref{section:leakage_quantification}.
The workflow extends the CellIFT methodology~\cite{solt2022cellift} to include leakage instrumentation for sequential cells, combinatorial cell handling, and counter insertion.

 \begin{figure*}[t]
    \centering
    \includegraphics[width=1\linewidth]{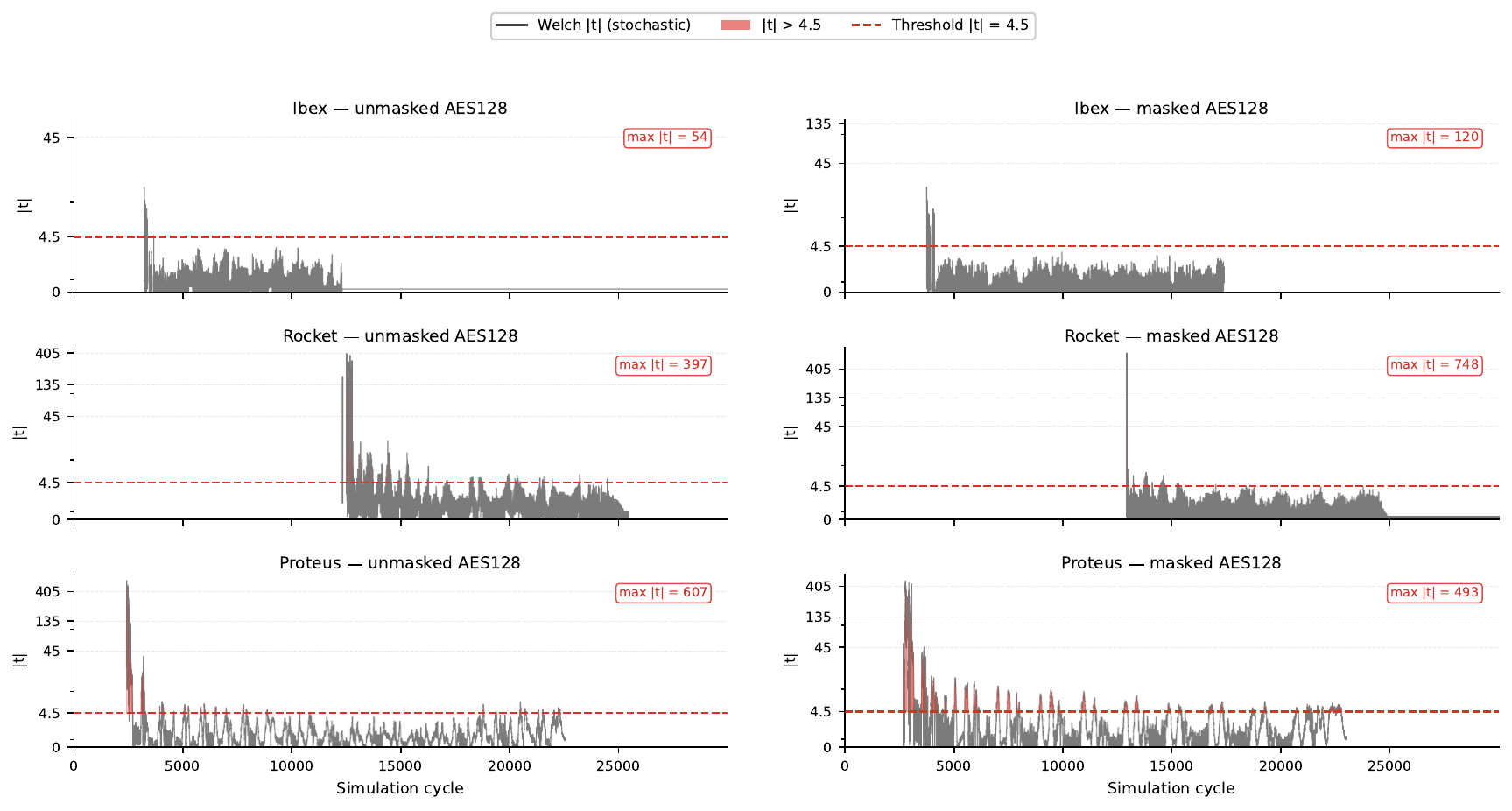}
    \caption{TVLA results for all processors running AES128}
    \Description{A three-by-two grid of line plots of Welch t-statistic magnitude against simulation cycle, from zero to roughly 27,000 cycles. Rows correspond to the Ibex, Rocket, and Proteus cores; the left column shows unmasked AES-128 and the right column masked AES-128. Each plot marks the TVLA threshold of 4.5 as a dashed horizontal line, with cycles exceeding it highlighted. All six configurations breach the threshold. Peak magnitudes are 54 and 120 for Ibex unmasked and masked, 397 and 748 for Rocket, and 607 and 493 for Proteus. Leakage appears in distinct intervals rather than uniformly across execution, and masking shifts the onset and shape of these intervals without eliminating them.}
    \label{fig:aes_leaky_cycles}
\end{figure*}

Following instrumentation, the design is compiled into a cycle-accurate simulation using Verilator \cite{verilator}. The HW and HD counters inserted by the Yosys pass are exposed as standard output ports in the simulation model, allowing sampling at every clock edge with minimal runtime overhead. This process produces a lightweight activity trace CSV file that records signal switching on leaky paths. For root-cause attribution, Verilator is configured to generate a single waveform in FST or VCD format for the entire workload. Although this waveform is comparatively large, only one simulation run is necessary to map the signals from the lightweight CSV trace and deduce the corresponding software instructions. Welch's t-test is computed per cycle across trace groups, and leakage intervals are identified by episode detection on the resulting t-statistic series. Instruction-level attribution maps program counter values recorded during simulation to source-level locations using the standard \texttt{addr2line} utility from the RISC-V GNU toolchain. The entire pipeline operates as a reproducible workflow, requiring no manual intervention beyond the initial configuration of the target processor and workload.

\subsubsection{Processors Under Evaluation
}
The methodology is evaluated on three RISC-V processors spanning a wide range of micro-architectural complexity, from a minimal embedded core to a full Linux-capable system. This selection enables assessment of the approach's generalizability across fundamentally different pipeline structures, memory hierarchies, and instruction-level parallelism strategies.\\
(i)~\texttt{Ibex} is a 2-stage in-order processor implementing the RV32IMC instruction set \cite{ibex}. Its minimal pipeline makes it representative of embedded cores, commonly used in security-critical applications, such as hardware roots of trust.\\
(ii)~\texttt{Proteus} is a design-time configurable RISC-V processor implemented in SpinalHDL, supporting the RV32IM instruction set \cite{bognar2023proteus}. It can be configured as either a classic five-stage in-order pipeline or a superscalar out-of-order processor with a configurable number of reorder-buffer entries and execution units. For this evaluation, Proteus is configured in out-of-order superscalar mode with eight arithmetic logic units (ALUs). This configuration of Proteus represents the high-performance end of the design spectrum.\\ 
(iii)~\texttt{Rocket} is a 5-stage in-order scalar processor implementing the RV64G instruction set, generated using the Chipyard SoC framework \cite{asanovic2016rocket, amid2020chipyard}. Rocket has branch prediction logic and a multi-level memory hierarchy, which introduces significantly more micro-architectural state and longer data propagation paths.

All processors are analyzed using bare-metal software binaries to ensure traceability and deterministic execution throughout the evaluation.

\subsubsection{Cryptographic Software Algorithms Under Evaluation}

Three software workloads are evaluated, each compiled to target the supported processor's ISA and executed in bare-metal mode. For Ibex, a 32-bit \texttt{march=rv32imc -mabi=ilp32} configuration is used, while for Proteus, a \texttt{march=rv32im -mabi=ilp32} configuration is applied; both utilize the \texttt{-O2} compiler flag. For Rocket (RV64G), the algorithms are compiled for the 64-bit datapath using the \texttt{-march=rv64g -mabi=lp64} configuration and the \texttt{-O2} compiler flag. All software is compiled with the RISC-V GNU toolchain version 12.2.0.
\\
(i)~\texttt{AES-128 (Unmasked)} implementation follows a standard byte-oriented AES-128 \cite{NIST-FIPS-197} structure, comprising \textit{AddRoundKey}, \textit{SubBytes} (table-based S-box lookup), \textit{ShiftRows}, and \textit{MixColumns} operations executed over ten rounds, with a key schedule that expands a 128-bit secret key into eleven round keys. No countermeasures are applied.\\
(ii)~\texttt{AES-128 (Masked)} implementation applies a first-order Boolean masking scheme to the AES-128 algorithm \cite{NIST-FIPS-197}. Each secret-derived intermediate value is split into two shares $v = v_{\text{data}} \oplus v_{\text{mask}}$, where $v_{\text{mask}}$ is drawn from a fresh random source at the start of each encryption to prevent mask reuse across invocations. The S-box is replaced with a masked variant that operates directly on shares, and \textit{MixColumns} is computed over the masked representation, with mask propagation maintained through all linear operations (\textit{ShiftRows}, \textit{AddRoundKey}). 
This countermeasure is theoretically sound at first order. Consequently, any leakage detected by the methodology indicates either an implementation flaw or a micro-architectural interaction that violates masking assumptions.\\
(iii)~\texttt{ML-KEM CRYSTALS-Kyber-512 (Unmasked)} is a decapsulation operation under the Kyber-512 parameter set, derived from CRYSTALS-Kyber, a quantum-resistant key encapsulation mechanism standardized by NIST~\cite{NIST-FIPS-203}. Only the unmasked variant is evaluated, providing a baseline for future masked Kyber studies. Kyber's polynomial arithmetic results in a structurally distinct leakage profile compared to AES: secrets appear in Number Theoretic Transform (NTT) coefficient arrays, with dominant operations such as integer multiplication, XOR, and shift replacing AES's byte-oriented S-box.

Evaluation of both AES and Kyber demonstrates the methodology's applicability across modern and post-quantum cryptographic algorithms. Assessment of both masked and unmasked AES further characterizes hardware implementation issues and the impact of CPU microarchitecture.

\subsubsection{Trace counts and simulation lengths} TVLA is applied per simulation cycle using Welch's t-test on matched-pair traces, where plaintext, ciphertext, and masking randomness are held constant across pairs and only the key differs, making the test sensitive exclusively to key-dependent leakage. For AES-128, 20 traces per group are used for the unmasked variant; this is increased to 100 traces per group for the masked variant to compensate for the reduced signal-to-noise ratio under masking. For Kyber-512, 20 traces per group are collected.

The AES and Kyber workloads differ substantially in computational demand. AES encryption completes in tens of thousands of cycles, whereas Kyber decapsulation requires millions of cycles. To enable a fair runtime comparison in SPARC, all AES simulations are executed for 30,000 cycles and all Kyber simulations for 2,000,000 cycles, irrespective of the processor.
Raw cycle counts do not directly reflect real-world performance. Proteus and Rocket feature multi-level cache hierarchies, while Ibex is connected directly to an SRAM model as main memory. This configuration does not represent realistic memory access behavior. However, since absolute performance is not the focus of these experiments, this distinction does not affect the validity of the results.
\begin{figure}[t]
    \centering
    \includegraphics[width=1\linewidth]{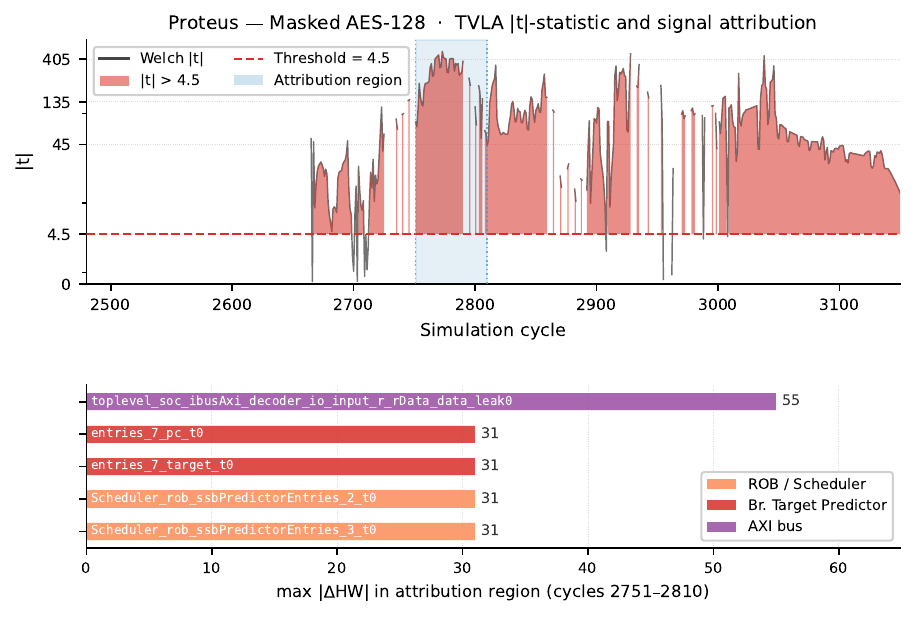}
    \caption{Signal Attribution for Proteus running masked AES128}
    \Description{Two stacked panels for Proteus running masked AES-128. The upper panel plots Welch t-statistic magnitude over simulation cycles 2500 to 3100, with values far exceeding the threshold of 4.5 and a shaded attribution region covering cycles 2751 to 2810. The lower panel is a horizontal bar chart ranking the five highest-contributing signals by maximum Hamming-weight difference within that region. An AXI bus read-data signal ranks highest at 55; four signals tie at 31, comprising two branch-target-predictor entries holding a program counter and a branch target, and two reorder-buffer scheduler predictor entries. Bars are colored by the microarchitectural structure they belong to: reorder buffer and scheduler, branch target predictor, and AXI bus.}
    \label{fig:proteus_singal_attribution}
\end{figure}
\subsection{Evaluation}
\label{sec:evaluation}
The evaluation addresses three research questions: the simulation cycles at which leakage occurs (RQ1), the signals contributing to leakage (RQ2), and the responsible software instructions (RQ3). Each question is examined across all three processors and both cryptographic workloads described in the experimental setup. For AES-128, both unmasked and first-order Boolean-masked variants are analyzed, enabling the methodology to characterize not only the presence of leakage but also the influence of masking on its onset and statistical strength. Kyber-512 is evaluated only in its unmasked form, offering a structurally distinct workload to assess generalization beyond block-cipher cryptography. Leakage is assessed per simulation cycle using TVLA with both HW and HD counters, applying a detection threshold of $|t|>4.5$.

\subsubsection{RQ1 - Leakage Detection: At Which Cycles Does the Design Leak?}

The first research question examines whether the taint-filtered TVLA methodology reliably detects key-dependent leakage and localizes it to specific simulation cycles.
All nine processor–algorithm combinations exceed the Welch’s t-test threshold at clock-cycle granularity, confirming that the instrumentation produces statistically significant leakage signals across all evaluated configurations. 
Figure~\ref{fig:aes_leaky_cycles} presents the per-cycle t-statistic time series for all processors executing AES, with the t-statistic threshold $|t| > 4.5$ indicated as a dashed line. Cycles exceeding this threshold are classified as leaky and are highlighted in the figure. The t-statistic trace reveals distinct leakage intervals corresponding to specific phases of the cryptographic computation. Leaky intervals are further investigated by signal and instruction attribution.

\subsubsection{RQ2 - Signal Attribution: Which Signals Contribute to the Leakage?}

While RQ1 identifies when leakage occurs, it does not indicate which signals contribute to the key-dependent switching. RQ2 addresses this by ranking individual leakage shadow signals within each detected leakage interval. As an example, we show Figure~\ref{fig:proteus_singal_attribution}, which displays the dominant attributed signals for Proteus running masked AES during the peak leakage interval identified in RQ1. For instance, we can observe that secret-dependent activity is exposed through the AXI bus signals, the reorder buffer, and the branch target predictor. This confirms the lack of a proper key-loading mechanism and indicates potential leakage due to microarchitectural artifacts in Proteus when running a software-masked AES-128.

\subsubsection{RQ3 - Instruction Attribution: Which Instructions Cause the Leakage?}
\begin{figure}[t]
    \centering
    \includegraphics[width=1\linewidth]{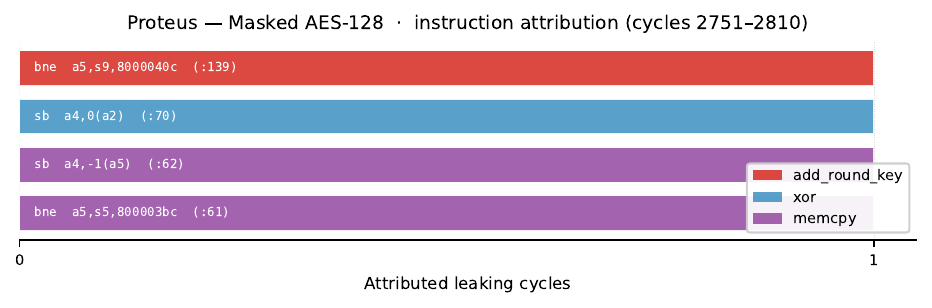}
    \caption{Instruction Attribution for Proteus running masked AES128}
    \Description{Horizontal bar chart attributing leaking cycles to individual instructions for Proteus running masked AES-128 over cycles 2751 to 2810. Four RISC-V instructions are listed with their source line numbers: two store-byte instructions and two branch-not-equal instructions. Bars are colored by the source-level function each instruction belongs to: add_round_key, xor, and memcpy. The horizontal axis gives the share of attributed leaking cycles.}
    \label{fig:proteus_instruction_attribution}
\end{figure}

RQ3 completes the root-cause analysis by linking the hardware leakage identified in RQ2 to the corresponding software instructions. In this step, SPARC identifies the dominant software instructions in the attribution region. Furthermore, the assembler instructions are mapped to the source core functions for a complete root-cause analysis. Figure~\ref{fig:proteus_instruction_attribution} presents the instruction-level attribution for Proteus running masked AES. Here, we can observe which exact load/store and branching instructions correspond to the attributed leaky hardware signals. During this interval, SPARC attributes multiple leaky signals to the four instructions shown in Fig.~\ref{fig:proteus_instruction_attribution}. Because a single instruction may excite several signals and out-of-order interleaving precludes a strict one-to-one correspondence, the complete active set is reported. Lastly, SPARC maps assembler instructions to functions in the original source code, such as \texttt{add\_round\_key}, \texttt{xor}, and \texttt{memcpy}.

\subsection{Framework Performance and Scalability}
\input{tables/framework_performance_comparison}
A key practical concern for any pre-silicon assessment methodology is whether it scales to realistic design sizes without huge runtime costs. We evaluate this by reporting instrumentation overhead, simulation time, and trace storage requirements across all three processors in Table \ref{tab:simtime}. All experiments are performed using an AMD Ryzen 7 7800X3D CPU with 64GB of DDR5 RAM. The instrumentation overhead covers the cost of performing the Yosys pass and building the executable Verilator simulations with compact CSV tracing. Inserting the HW/HD counters enlarges the instrumented design by $\sim$11-12\% over pure cell-level IFT; counters and shadow logic exist only in the evaluation model. Even for more complex processors like Rocket and Proteus, the overhead remains reasonable, requiring less than 4 and 9 minutes, respectively. The compact CSV trace simulator, which produces quantified leakage traces, is very fast,
achieving an 8x speedup in per-trace simulation time on Ibex AES relative to Telescopes' \cite{liu2025telescope} reported simulation cost for a comparable design. This scalability is a direct result of taint filtering. For Ibex, approximately 31\% of all design signals are located on a tainted path, and only about 16\% both reside on a leaky path and switch during execution. In contrast, approaches that do not use IFT must process the switching activity of every signal. The runtime scales well with larger processors and enables practical evaluation of workloads that span millions of cycles, such as Kyber. The total evaluation time encompasses several steps: building all simulation targets, including the VCD/FST dump simulator required for the root-cause analysis; Running all the defined test-patterns with the CSV trace simulator; Running a single simulation with the complete VCD/FST dump; Performing TVLA based on the obtained traces and attributing the hardware signals and software instructions in leaky intervals based on the complete VCD/FST simulation dump. The full evaluation and root-cause analysis of AES (both masked and unmasked) take less than 25 minutes on the most complex processor, while Kyber can be evaluated in just over 10 hours. 

Similar to other dynamic analyses, paths that are not activated by the selected inputs and keys remain unassessed. Additionally, effects occurring below the macro-cell abstraction, such as glitches resulting from technology mapping, are intentionally excluded from the model.

%% file: tables/framework_performance_comparison.tex
%ibex:eval:aes — 1m 25s (85s total)
%proteus:eval:aes — 24m 34s (1474s total)
%rocket:eval:aes — 13m 45s (825s total)

\begin{table}[t]
\centering
\footnotesize
\caption{Per-trace simulation time and total evaluation time across processors and workloads}
\label{tab:simtime}
\begin{tabularx}{\columnwidth}{@{}lXrrrr@{}}
\toprule
\textbf{Core} & \textbf{Workload} & \textbf{Cycles} & \textbf{\makecell[r]{Instrumentation}} & \textbf{\makecell[r]{Single CSV\\Trace}} & \textbf{\makecell[r]{Total\\Time}} \\
\midrule
Ibex    & AES$^*$   & 30k & 9s & 0.9s     & 1m 25s   \\
Rocket  & AES$^*$   & 30k & 3m 12s & 17.1s  & 13m 45s  \\
Proteus & AES$^*$   & 30k & 8m 40s & 18.2s    & 24m 34s  \\
Ibex    & Kyber-512 & 2M  & 9s & 1m 6s    & 13m 34s  \\
Rocket  & Kyber-512 & 2M  & 3m 12s & 18m 49s  & 4h 14m   \\
Proteus & Kyber-512 & 2M  & 8m 40s & 22m 14s & 10h 4m   \\
\bottomrule
\end{tabularx}
\par\smallskip
{\footnotesize $^*$ Total time includes both masked and unmasked evaluation.}
\end{table}

%% file: sections/06-conclusion.tex
\section{Conclusion}
This work presents SPARC, an automated framework for pre-silicon PSCL evaluation and root cause analysis in the processor design flow. By testing three diverse RISC-V processors: Ibex (2-stage in-order, 32-bit), Proteus (out-of-order superscalar, 32-bit), and Rocket (5-stage in-order, 64-bit) and evaluating two distinct cryptographic algorithms (AES-128 and CRYSTALS-Kyber-512), including a first-order Boolean-masked AES-128 implementation, we demonstrate the efficacy and scalability of SPARC. Our experiments confirm the well-known power side-channel leakage points in unmasked implementations and, critically, detect micro-architectural leakage artifacts in Proteus's out-of-order pipeline that violate the masking assumptions and expose key-dependent information despite first-order protection. Furthermore, SPARC goes beyond mere detection: it pinpoints the exact root cause, identifying the specific hardware signals, micro-architectural artifacts, and executing software functions responsible for the leak. Our trace generation outperforms previously reported solutions by a factor of 8. With a total evaluation runtime of less than 25 minutes for AES-128 and around 10 hours for Kyber-512 for Proteus, SPARC delivers an effective and scalable framework for pre-silicon PSCL evaluation and root cause analysis. Guided by the root-cause output, designers can revise the RTL, such as by implementing a dedicated key-loading path or clearing secret-holding states, and then immediately re-evaluate the design. Alternatively, the ranked signals may be flagged as hardware-masking candidates for subsequent design stages. Future work includes gate-level cross-validation of attribution accuracy, implementation of masked ML-KEM, and exploration of multi-core SoCs.